\documentclass{emulateapj}
\usepackage{apjfonts}
\usepackage{graphicx}
\usepackage{amsmath}
     \usepackage{dpfloat}

\usepackage{longtable}

\usepackage{color}

\def\kms{\ifmmode{\rm km\thinspace s^{-1}}\else km\thinspace s$^{-1}$\fi}

\shortauthors{Sanchis-Ojeda et al.~2015}
\shorttitle{RM effect of WASP-47b}

\begin{document}

%
\def\ltsima{$\; \buildrel < \over \sim \;$}
\def\lsim{\lower.5ex\hbox{\ltsima}}
\def\gtsima{$\; \buildrel > \over \sim \;$}
\def\gsim{\lower.5ex\hbox{\gtsima}}
%

\bibliographystyle{apj}

\title{A low stellar obliquity for WASP-47, a compact multiplanet system \\
with a hot Jupiter and an ultra-short period planet}

\author{
Roberto~Sanchis-Ojeda\altaffilmark{1, 2},
Joshua~N.~Winn\altaffilmark{3}, 
Fei~Dai\altaffilmark{3},
Andrew~W.~Howard\altaffilmark{4},
Howard~Isaacson\altaffilmark{1}, \\
Geoffrey~W.~Marcy\altaffilmark{1},
Erik~Petigura\altaffilmark{5,6},
Evan~Sinukoff\altaffilmark{4}, 
Lauren~Weiss\altaffilmark{1, 7}, \\ 
Simon~Albrecht\altaffilmark{8},
Teruyuki~Hirano\altaffilmark{9}, 
Leslie~Rogers\altaffilmark{5,6}
}

\altaffiltext{1}{Department of Astronomy, University of California, Berkeley, CA 94720; sanchisojeda@berkeley.edu}

\altaffiltext{2}{NASA Sagan Fellow}

\altaffiltext{3}{Department of Physics, and Kavli Institute for
  Astrophysics and Space Research, Massachusetts Institute of
  Technology, Cambridge, MA 02139, USA}

\altaffiltext{4}{Institute for Astronomy, University of Hawaii, 2680 Woodlawn Drive, Honolulu, HI 96822, USA}

\altaffiltext{5}{Department of Astronomy and Division of Geological and Planetary Sciences, California Institute of Technology, Pasadena, CA 91125, USA}

\altaffiltext{6}{NASA Hubble Fellow}

\altaffiltext{7}{Ken \& Gloria Levy Fellow}

\altaffiltext{8}{Stellar Astrophysics Centre, Department of Physics and Astronomy, Aarhus University, Ny Munkegade 120, DK-8000 Aarhus C, Denmark}

\altaffiltext{9}{Department of Earth and Planetary Sciences, Tokyo Institute of Technology, 2-12-1 Ookayama, Meguro-ku, Tokyo 152-8551, Japan}

\slugcomment{Submitted to the {\it Astrophysical Journal Letters}, 2015 August 27}

\begin{abstract}

  We have detected the Rossiter-Mclaughlin effect during a transit of WASP-47b, the only known hot Jupiter with close planetary companions. By combining our spectroscopic observations with {\it Kepler} photometry, we show that the projected stellar obliquity is $\lambda = 0^\circ \pm 24^\circ$.  We can firmly exclude a retrograde orbit for WASP-47b, and rule out strongly misaligned prograde orbits. Low obliquities have also been found for most of the other compact multiplanet systems that have been investigated.  The Kepler-56 system, with two close-in gas giants transiting their subgiant host star with an obliquity of at least 45$^\circ$, remains the only clear counterexample.

\end{abstract}


\section{Introduction}

Stars that host close-in giant planets have been found to have a wide range of obliquities (see, e.g., Triaud et al.\ 2010, Albrecht et al.\ 2012, or the review by Winn \& Fabrycky 2015). Attempts to explain this surprising result have until recently focused on the implications for the formation mechanism of hot Jupiters. The broad range of obliquities seemed to suggest that whichever processes shrink the orbits of hot Jupiters are also capable of tilting their orbits away from the plane of formation. It remains possible, though, that the obliquity distribution has nothing to do with hot Jupiters {\it per se}, but is instead the outcome of more general post-formation dynamical processes.

This ambiguity arises because measurements of stellar obliquity have been largely confined to hot Jupiters, for practical reasons.  The main technique used to measure obliquities is the Rossiter-McLaughlin (RM) effect, which is easier to observe for larger planets in short-period orbits. More recently it has become possible to probe the obliquities of other types of systems, particularly the compact systems of multiple transiting planets found using data from the {\it Kepler} spacecraft. The RM effect is more difficult to observe in such systems because the planets are generally smaller and the transits are less frequent than for hot Jupiters, but detections have been achieved for two such systems: Kepler-25 and Kepler-89 (Hirano et al.\ 2012, Albrecht et al.\ 2013). Both stars have low obliquities.  Low obliquities were also measured for the Kepler-30 system by analyzing starspot-crossing events (Sanchis-Ojeda et al.\ 2012) and for the Kepler-50 and Kepler-65 systems by analyzing the rotational splittings of $p$-mode oscillations (Chaplin et al.\ 2013). Thus it seemed that the stellar hosts of compact multiplanet systems generally have low obliquities (Albrecht et al.\ 2013), and that the high obliquities were indeed peculiar to hot Jupiters.

This tentative conclusion was shattered by the discovery that the host star of Kepler-56, a compact system of two transiting planets, has an obliquity of at least $45^\circ$ (Huber et al.\ 2013). Batygin (2012) anticipated the existence of such systems, as the outcome of disk migration in the presence of a distant companion star on a misaligned orbit.  Li et al.\ (2015) demonstrated that the misalignment of Kepler-56 could have also arisen independently of disk migration, as a consequence of planet-planet dynamics. Thus, it now seems possible that the obliquities of hot-Jupiter hosts gave us a glimpse of more widespread phenomena, and that we might expect a wide range of stellar obliquities for planetary systems in general.

In this light, the WASP-47 system has a special importance, as the only known case of a hot Jupiter that is also part of a compact multiplanet system. The hot Jupiter in this ``hybrid'' system has an orbital period of $P = 4.1$~days and was discovered several years ago (Hellier et al.\ 2012).  Subsequently, long-term radial-velocity monitoring revealed a Jovian planet with a much longer period (Neveau-VanMalle et al. 2015, submitted). Most recently, data from the {\it K2} mission (Howell et al.\ 2014) revealed that there are two additional transiting planets, a super-Earth with a period of 0.79 days and a Neptune-size planet in a 9-day orbit (Becker et al.\ 2015).  Such configurations are thought to be rare (Steffen et al.\ 2012).  Measuring the obliquities of the stars in such systems might help to determine the relationship between them, the other compact multiple-planet systems, and the more isolated hot Jupiters.

This paper is organized as follows. Section~\ref{sec:obs} presents the spectroscopic and photometric observations of WASP-47 that were used in our study. Section~\ref{sec:analysis} presents our analysis leading to the determination of the projected stellar obliquity.  Section~\ref{sec:discussion} discusses these results and compares WASP-47 to other known planetary systems.

\section{Observations and data reduction}
\label{sec:obs}

WASP-47 happened to fall within the field of view of ``Campaign 3'' of the {\it K2} mission, which uses the {\it Kepler} space telescope to perform precise photometry of stars in selected ecliptic fields (Howell et al.\ 2014).  Becker et al.\ (2015) have presented an analysis of the photometric data, including the discovery of the additional planets ``d'' and ``e''.  Here we use the short-cadence photometric time series provided by Becker et al.\ (2015), which have a time sampling of one minute and were corrected for long-term trends due to stellar variability and instrumental artifacts.

In an attempt to detect the RM effect, we monitored the optical spectrum at high resolution using the HIRES spectrograph (Vogt et al.\ 1994) on the Keck~I telescope, on the night of UT August 10, 2015.  The weather was excellent, with subarcsecond seeing. We obtained 29 spectra over the course of 5~hours spanning a transit of the hot Jupiter WASP-47b.  Exposure times were 11-13 minutes, set by an exposure meter that enforced a signal-to-noise (S/N) ratio of 100~pixel$^{-1}$ near 550~nm for each spectrum. We used the 0.87$''$-wide C2 decker giving a spectral resolution of 55,000.  A glass cell of molecular iodine was placed in the light path to calibrate the HIRES spectra.  The next night, we obtained a spectrum of WASP-47 without the iodine cell, and with S/N = 200~pixel$^{-1}$, to serve as the template spectrum in the Doppler analysis.

The reduction of HIRES spectra and the Doppler analysis used the standard pipeline of the California Planet Survey (Howard et al.\ 2010). The deconvolved iodine-free spectrum of WASP-47 was used with the molecular iodine line atlas in a forward model that determines the wavelength zero point, spectral dispersion, Doppler shift, and the instrumental profile, following the methods of Marcy \& Butler (1992). The forward model works on 2~\AA~spectral segments and uses the weighted average to determine the final radial velocity for each of the observations. The results are shown in Figure~\ref{fig:RM} and presented in Table~\ref{tbl:rvs}.

\begin{deluxetable}{ccc}
\tabletypesize{\scriptsize}
\tablecaption{Radial velocity observations\label{tbl:rvs}}
\tablewidth{0pt}

\tablehead{
\colhead{Time [BJD]} & \colhead{Radial velocity [m/s]} & \colhead{Uncertainty [m/s]} 
}

\startdata
 2457244.871873 &       24.1 &        1.4 \\ 
 2457244.879755 &       29.7 &        1.6 \\ 
 2457244.887822 &       22.8 &        1.5 \\ 
 2457244.896063 &       19.5 &        1.7 \\ 
 2457244.904292 &       20.7 &        1.6 \\ 
 2457244.912684 &       20.0 &        1.6 \\ 
 2457244.920959 &       17.6 &        1.5 \\ 
 2457244.929316 &       16.2 &        1.7 \\ 
 2457244.937406 &       14.6 &        1.6 \\ 
 2457244.945624 &       20.8 &        1.6 \\ 
 2457244.953888 &       22.4 &        1.5 \\ 
 2457244.962233 &       22.6 &        1.6 \\ 
 2457244.970729 &       18.9 &        1.6 \\ 
 2457244.979537 &       12.9 &        1.6 \\ 
 2457244.988021 &        4.2 &        1.6 \\ 
 2457244.996621 &        0.7 &        1.6 \\ 
 2457245.005799 &       -6.0 &        1.7 \\ 
 2457245.014526 &       -9.0 &        1.6 \\ 
 2457245.023033 &      -16.5 &        1.5 \\ 
 2457245.031413 &      -17.7 &        1.6 \\ 
 2457245.039723 &      -20.2 &        1.5 \\ 
 2457245.048254 &      -25.4 &        1.6 \\ 
 2457245.057004 &      -29.9 &        1.5 \\ 
 2457245.066506 &      -29.9 &        1.8 \\ 
 2457245.075314 &      -24.3 &        1.5 \\ 
 2457245.083902 &      -21.3 &        1.5 \\ 
 2457245.092618 &      -24.9 &        1.5 \\ 
 2457245.101518 &      -33.1 &        1.5 \\ 
 2457245.110824 &      -29.9 &        1.5 
 \enddata
\end{deluxetable}

\section{Analysis of the observations}
\label{sec:analysis}

\subsection{Determination of projected rotation rate}

The transits of WASP-47b are very nearly central, i.e., the transit impact parameter $b$ is close to zero. In such situations, analyses of the RM effect are subject to a strong degeneracy between the sky-projected stellar obliquity $\lambda$ and the sky-projected stellar rotation rate $v\sin{i}$ (Gaudi \& Winn 2007). Therefore it is of special importance to determine or constrain the value of $v\sin{i}$ independently of the RM effect.

We analyzed the iodine-free, high-S/N spectrum using {\tt SpecMatch} (Petigura 2015). This is a code for obtaining the basic spectroscopic parameters from high-resolution spectra. It creates synthetic spectra for any input values of $T_{\rm eff}$, $\log g$, and [Fe/H], by interpolating between a grid of model spectra taken from the library of Coelho et al.\ (2005).  {\tt SpecMatch} also applies broadening kernels to account for rotational and macro-turbulent broadening. Importantly, {\tt SpecMatch} also measures the HIRES instrumental
line broadening, which depends on the slit width and atmospheric seeing at the time of observations, by fitting the $O_2$ B-X telluric lines with a comb of Gaussians. For WASP-47 the instrumental profile is modeled as a Gaussian function with a dispersion of 1.95~km~s$^{-1}$.

Given an observed spectrum, {\tt SpecMatch} adjusts the spectroscopic parameters of the synthetic spectrum until it obtains the best match.  For WASP-47, the results were $T_{\rm eff} = 5565 \pm 60$~K, $\log g = 4.29\pm 0.07$, and [Fe/H] = $0.43 \pm 0.04$. These values are all in agreement with those presented by Hellier et al.\ (2012) and Mortier et al.\ (2013).  For the projected rotation rate, {\tt SpecMatch} found an upper bound $v \sin{i} = 0 \pm 2$~km~s$^{-1}$, based on the lack of detectable rotational broadening. This result is slightly different from the value of $3.0 \pm 0.6$~km~s$^{-1}$ reported by Hellier et al.\ (2012). To check on the {\tt SpecMatch} result, we also performed an independent analysis of the line profile using the tools described by Hirano et al.\ (2014). We found $v \sin{i} = 1.3^{+1.0}_{-1.2}$~km~s$^{-1}$, in good agreement with the value found by {\tt SpecMatch}. In the subsequent analysis we use $v \sin{i} = 0 \pm 2$~km~s$^{-1}$, because it is based on a higher-S/N and higher-resolution spectrum, and because we have calibrated the $v\sin{i}$ scale of {\tt SpecMatch} against all our previous RM results. We checked that this choice did not affect our final conclusions significantly, by using the other two constraints, obtaining similar results.

\begin{figure}[h]
\begin{center}
\includegraphics[width=0.485 \textwidth]{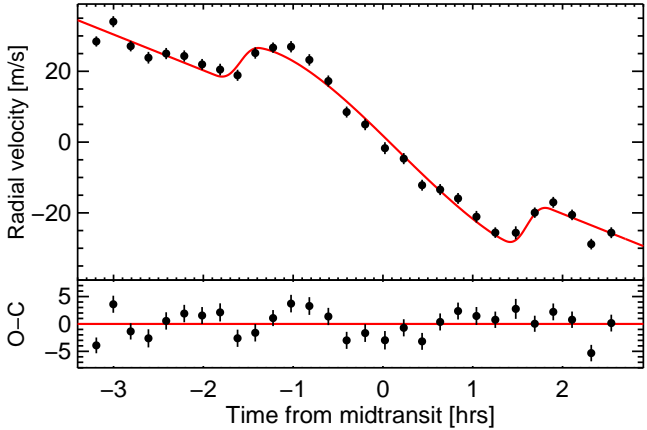} 
\caption{
{\bf RM model fit to the radial velocities during a transit of WASP-47b.}
Radial velocity observations are represented with black dots, with uncertainty bars that are sometimes smaller than the dots (and do not include the jitter term). The best-fit RM model to the data shown in the upper panel gives a $\lambda$ near zero, and it is represented with a red thick line. The lower panel shows the residual radial velocity respect to the model. 
}
\label{fig:RM}
\vspace{-0.1cm}
\end{center}
\end{figure}

\subsection{Determination of the projected obliquity}

Our first goal was to create a phase-folded light curve for the transits of WASP-47b, which could be fitted simultaneously with the Doppler data.  For this purpose we used the detrended photometric time series provided by Becker et al.\ (2015).  We generated an initial phase-folded transit light curve for WASP-47b, using the ephemerides of Becker et al.\ (2015) to fold the data with the constant period of WASP-47b and discarding the data that were obtained during transits of planets d and e.

Next, we wanted to improve on this light curve by taking into account the transit-timing variations of WASP-47b.  We fitted the initial light curve with the model of Mandel \& Agol (2002) for the case of a quadratic limb-darkening law. The free parameters were the transit depth $(R_{\rm p}/R_\star)^2$, impact parameter $b$, the scaled stellar radius $R_\star/a$, and the limb-darkening coefficients $u_1$ and $u_2$. Then the best-fitting parameters were held fixed, and the model was used as a template to determine the midtransit time of each individual transit. We fitted the 5 hours of data spanning each transit with 3 free parameters: the midtransit time, and the two parameters describing a linear function of time (to take into account the covariance between the midtransit time and longer-term trends in stellar brightness). The series of midtransit times determined in this manner were well-fitted by a quadratic function of transit epoch, which takes into account the observed transit timing variations that have a maximum deviation of 1 minute\footnote{Becker et al.\ (2015) modeled the timing variations as a sinusoidal function of epoch, which gives essentially the same results.}. This new orbital ephemeris was then used to re-fold the original time series, producing a new phase-folded light curve for WASP-47b that takes into account the small transit-timing variations.  To speed subsequent computations we averaged the light curve into one-minute bins. The resulting light curve is well-fitted by a standard transit model, with residuals that appear random and independent. To determine appropriate weights for the binned data points, we assigned an uncertainty to each point such that $\chi^2_{\rm min} = N_{\rm dof}$.

We then fitted simultaneously the radial-velocity data and the binned photometric data.  To describe the photometric transit we used the same Mandel \& Agol (2002) model with the same five free parameters, but with both limb-darkening coefficients subject to Gaussian priors based on theoretical estimates (Claret \& Bloemen 2011). The mid-transit time of the folded light curve was allowed to vary freely.

To describe the RM effect, we used the code of Hirano et al.\ (2011). This code takes into account all the important line-broadening mechanisms as well as the RM effect. The key parameters were $v\sin i$ and $\lambda$.  For $v \sin{i}$, we required it to be positive, and used a Gaussian prior with a mean of zero and a dispersion of 2~km~s$^{-1}$, based on our findings from the previous section.  We also modeled the out-of-transit radial-velocity variation with a linear function of time, giving two additional parameters $\gamma$ and $\dot{\gamma}$. We adopted a Lorentzian (natural) broadening of 1~km~s$^{-1}$ and a Gaussian (instrumental plus thermal) broadening of 3.3~km~s$^{-1}$, based on the instrumental profile of HIRES and the effective temperature of the star.  The macro-turbulent broadening was allowed to vary, subject to Gaussian prior corresponding to $3\pm 0.5$~km~s$^{-1}$, based on the relationships given by Doyle et al.\ (2014).  The convective blueshift was taken into account with an assumed amplitude of $0.5$~km~s$^{-1}$, following Albrecht et al.\ (2012). We decided to allow the midtransit time to vary freely, rather than constraining it to conform to the transit ephemeris, given the observed level of transit-timing variations in the system.

The last parameter of the model is the ``stellar jitter'' term, which
is added in quadrature to the radial-velocity uncertainties that are
internally estimated by our Doppler code. We decide to allow the jitter term to vary freely, so rather than minimizing the $\chi^2$ function, we minimized the negative of the logarithm of the likelihood. In this way we can naturally control the jitter term by adding the log of the final radial velocity uncertainty (see equation 1 of Johnson et al. 2011). This ensures a similar result to what we would find fixing the jitter term requiring the best-fit $\chi^2 = N_{\rm dof}$, 

We determined the allowed ranges for the 13 model parameters, and their covariances, with a Monte Carlo Markov Chain routine.  Figure~\ref{fig:RM} shows the best-fitting model to the RV signal.  The results for the marginalized distribution of each relevant parameter can be found in Table~\ref{tbl:params}. Our photometric parameters are all compatible with those found by Becker et al.\ (2015), including the midtransit time. The stellar jitter of 2.1~m~s$^{-1}$ is a typical value for a star of this type. From the value of $\gamma$ we can get an estimate of the radial velocity semi-amplitude K of $K_{\rm RM} = -\dot{\gamma}P_{orb}/2\pi = 161 \pm 5$~m~s$^{-1}$, slightly larger than the value of $K = 136 \pm 5$~m~s$^{-1}$ found in the discovery paper (Hellier et al. 2012). 

\begin{deluxetable}{lcc}
\tabletypesize{\scriptsize}
\tablecaption{Key parameters of WASP-47b\label{tbl:params}}
\tablewidth{0pt}

\tablehead{
\colhead{Parameter} & \colhead{Value} & \colhead{68.3\% Conf.~Limits} 
}

\startdata
Sky-project obliquity, $\lambda$ [deg] & 0 & $\pm 24$ \\
$v \sin{i}$ [km~s$^{-1}$] &  1.80 & $ (+0.24, -0.16)$   \\
RV offset, $\gamma$ [m~s$^{-1}$] & $-4.3$  & $\pm 0.7$ \\
RV slope, $\dot{\gamma}$ [m~s$^{-1}$~day$^{-1}$] & $-244$ & $\pm 8$ \\
Stellar jitter, $\sigma$ [m~s$^{-1}$] & $2.1$ & $\pm 0.5$ \\
Midtransit time [BJD$_{\rm TDB}$]  & $ 2457245.0049$  &  $\pm 0.0017$      \\
$(R_{\textrm{p}}/R_\star)^2 $  & $0.01034$         &  $\pm 0.00004$  \\
$R_\star/a$      & $0.1025$         &  $\pm 0.0007$  \\
Impact parameter, $b$ & 0.13  & $(+0.05, -0.08)$ 
\enddata
\end{deluxetable}

\begin{figure}[h]
\begin{center}
\includegraphics[width=0.485 \textwidth]{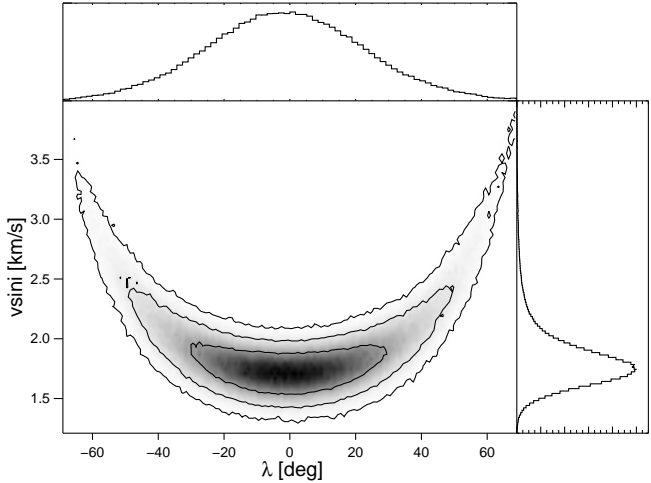} 
\caption{
{\bf The degeneracy between $\lambda$ and $v \sin{i}$ increases the uncertainty on $\lambda$.}
The central plot shows a contour plot for the combined distribution of values for $\lambda$ and $v \sin{i}$ from our MCMC routine. The three solid lines represent the 1, 2 and 3$\sigma$ contours. Darker regions represent areas with higher posterior density. A better knowledge of $v \sin{i}$ (beyond our upper bound of 2~km~s$^{-1}$) would allow a better determination of $\lambda$.}
\label{fig:correlation}
\end{center}
\end{figure}

As expected the uncertainty in $\lambda$ is dominated by the covariance with $v\sin i$, a characteristic of systems with a low impact parameter. Figure~\ref{fig:correlation} shows the joint posterior distribution function of $\lambda$ and $v \sin{i}$ resulting from our MCMC analysis.  The covariance between $\lambda$ and $v \sin{i}$ causes the uncertainty in $\lambda$ to be larger than one might expect from such a high-S/N detection of the RM effect.  If it were possible to determine $v \sin{i}$ accurately and precisely from external observations, this degeneracy could be broken, and a precise value of $\lambda$ could be obtained, but this is not yet the case for WASP-47.

\section{Discussion}
\label{sec:discussion}

We have presented the third measurement of the RM effect for a compact multiplanet system. In all three cases the projected obliquity has been found to be consistent with zero. For WASP-47 our determination of $\lambda$ was limited by the strong covariance between $\lambda$ and $v \sin{i}$, an effect caused by the low impact parameter of the planetary transits. In principle this problem could be ameliorated by a better determination of the impact parameter, but this is unlikely to be achieved soon, given that the {\it K2} data already has such high photometric precision.  Another way to improve the measurement of $\lambda$ would be to refine our constraint on $v \sin{i}$, either via new and higher-resolution spectroscopic observations, or by measuring the rotation period of the host star. Regarding the latter, Hellier et al.\ (2012) did not report any quasiperiodic variability in the WASP data that could be attributed to stellar rotation, and likewise, we have not found any convincing signals in the {\it K2} data.

Our results have allowed us to exclude a retrograde orbit for WASP-47b, and to rule out strongly misaligned prograde orbits. As recently as 2008, this would not have been the least bit surprising.  In particular, the theory of disk migration --- a leading contender to explain the existence of close-in planets and multiplanet systems --- was expected to lead to well-aligned and low-obliquity systems. But in the past few years we have learned that stellar obliquities can vary widely, for reasons that are not yet understood. For WASP-47, a low obliquity or even a moderate obliquity was not a foregone conclusion, especially given how unusual the system is in comparison to the other known systems.

Consider, for example, a comparison between WASP-47 and other systems discovered by {\it Kepler} with at least one transiting planet with a period shorter than one day (the ``ultra-short-period'' planets studied by Sanchis-Ojeda et al.\ 2014). The {\it Kepler} telescope has revealed additional transiting companions in these systems, but these companions are typically Neptune-sized or smaller. In only one case, KOI~191, is there a Jupiter-sized companion planet similar to WASP-47b, and in that system the giant planet has a period of 15 days.

Or consider the class of systems with one close-in giant planet and at least one additional planet. Using the online planet catalog {\tt exoplanets.org} (Han et al.\ 2014), we found that among the 205 planetary systems with a transiting planet larger than 0.5~$R_{\rm Jup}$ and orbital periods shorter than 15 days, only 5 of them have known planetary companions.  Some of these are Jupiter-size planets with very long-period companions (see, e.g., Knutson et al.\ 2014). Others have planets that are half the size of Jupiter and have smaller, shorter-period companions (such as Kepler-18, Cochran et al.\ 2011). None of them resemble WASP-47, which has a radius of 1.1~$R_{\rm Jup}$, as well as nearby companions both interior and exterior.

Finally, consider a comparison with a mass-selected sample. There are 245 known stars with at least one planet with a mass larger than 0.1~$M_{\rm Jup}$ and a period shorter than 15~days.  Only 11 of these have more than one known planet, many of which were also in the radius-selected sample described above. The only system within this sample that resembles WASP-47 in having a close-in giant planet with a small interior companion and at least one exterior companion is 55~Cnc (Butler et al.\ 1997, Marcy et al.\ 2002, Fischer et al.\ 2008, Winn et al.\ 2011).  This planetary system has a giant planet with an orbital period of 14.6 days, and a 2~$R_E$ planet with a period of 17.7 hours, along with a few other gas giants with periods of 44, 260 and 5170 days. Thus, WASP-47 might be considered a more compact cousin of the 55~Cnc system. In this respect it is interesting to note that Bourrier \& H\'ebrard (2014) found 55~Cnc to have a projected obliquity of 72$^\circ$, although this conclusion has been disputed by Lopez-Morales et al.\ (2014).

We would need to extend the orbital period range of the transiting Jupiter-size planet up to 25 days to find another example of a system with a planet smaller than Neptune interior to the orbit of a warm Jupiter-size planet. The Kepler-89 system (Weiss et al. 2013) contains a mini-Neptune planet with an orbital period of 10 days, a gas giant with a period of 22.3 days and an inner smaller planet with an orbital period of 3.7 days. In this case there are two planets interior to the orbit of the Jupiter-size planet, but they are not as close to the host star as the super-earths 55~Cnc e and WASP-47d. 

It remains the case that the only clear and undisputed example of a compact multiplanet system with a high stellar obliquity is Kepler-56 (Huber et al.\ 2013). Additional measurements of obliquities in multiplanet systems, either in unusual systems such as WASP-47 or using other techniques more capable to study larger samples of diverse planet hosts (Morton \& Winn 2014; Mazeh et al. 2015), may play an important role in future attempts to understand their formation and their relationship to more isolated hot-Jupiter systems.

\acknowledgements We thank Eugene Chiang and Rebekah Dawson for helpful discussions. We are grateful to Juliette Becker and Andrew Vanderburg for sharing their manuscript at an early stage and for making their data publicly available. This work was performed, in part, under contract with the Jet Propulsion Laboratory (JPL) funded by NASA through the Sagan Fellowship Program executed by the NASA Exoplanet Science Institute. This research has made use of the Exoplanet Orbit Database and the Exoplanet Data Explorer at exoplanets.org. This work was supported by a NASA Keck PI Data Award, administered by the NASA Exoplanet Science Institute. Work by JNW was supported by the NASA Origins program (grant code NNX11AG85G). LMW thanks Ken and Gloria Levy for their generous support. T.H. is supported by Japan Society for Promotion of Science (JSPS) Fellowship for Research (No.25-3183). LAR gratefully acknowledges support provided by NASA through Hubble Fellowship grant \#HF-51313 awarded by the Space Telescope Science Institute, which is operated by the Association of Universities for Research in Astronomy, Inc., for NASA, under contract NAS 5-26555. Data presented herein were obtained at the W.M.~Keck Observatory from telescope time allocated to the National Aeronautics and Space Administration through the agency's scientific partnership with the California Institute of Technology and the University of California. The Observatory was made possible by the generous financial support of the W.M.~Keck Foundation. The authors wish to recognize and acknowledge the very significant cultural role and reverence that the summit of Mauna Kea has always had within the indigenous Hawaiian community. We are most fortunate to have the opportunity to conduct observations from this mountain.

\end{document}